\documentclass[pdflatex,sn-mathphys-num]{sn-jnl}% Math and Physical Sciences Numbered Reference Style 
\usepackage{graphicx}%
\usepackage{multirow}%
\usepackage{amsmath,amssymb,amsfonts}%
\usepackage{amsthm}%
\usepackage{mathrsfs}%
\usepackage[title]{appendix}%
\usepackage{xcolor}%
\usepackage{textcomp}%
\usepackage{manyfoot}%
\usepackage{booktabs}%
\usepackage{algorithm}%
\usepackage{algorithmicx}%
\usepackage{algpseudocode}%
\usepackage{listings}%
\usepackage{subfigure}
\theoremstyle{thmstyleone}%
\theoremstyle{thmstyletwo}%

\theoremstyle{thmstylethree}%

\begin{document}

\title[Article Title]{An Integrated Deep Learning Model for Skin
Cancer Detection Using Hybrid Feature Fusion
Technique}


\author[1]{\fnm{Maksuda Akter} \sur{Akter}}\email{maksudaoni6@gmail.com}

\author[1,2]{\fnm{Rabea} \sur{Khatun}}\email{rabeakhatun650@gmail.com}

\author[3]{\fnm{Md. Alamin} \sur{Talukder}}\email{alamin.cse@iubat.edu}
\author[1]{\fnm{Md. Manowarul } \sur{Islam}}\email{manowar@cse.jnu.ac.bd}
\author*[1]{\fnm{Dr. Md. Ashraf } \sur{ Uddin}}\email{ashraf@cse.jnu.ac.bd}

\affil*[1]{\orgdiv{Department of Computer Science and Engineering}, \orgname{Jagannath University}, \orgaddress{\street{9-10 Chittaranjan Ave}, \city{Dhaka}, \postcode{1100},\state{} \country{Bangladesh}}}

\affil[2]{\orgdiv{Department of Computer Science and Engineering}, \orgname{Green University of Bangladesh}, \orgaddress{\street{Purbachal American City}, \city{Kanchan, Rupganj, Narayanganj}, \postcode{1461}, \state{Dhaka}, \country{Bangladesh}}}

\affil[3]{\orgdiv{Department of Computer Science and Engineering}, \orgname{International University of Business Agriculture and Technology}, \orgaddress{\state{Dhaka}, \country{Bangladesh}}}

\abstract{Skin cancer is a serious and potentially fatal disease caused by DNA damage. Early detection significantly increases survival rates, making accurate diagnosis crucial. In this
groundbreaking study, we present a hybrid framework based
on Deep Learning (DL) that achieves precise classification of
benign and malignant skin lesions. Our approach begins with
dataset preprocessing to enhance classification accuracy, followed
by training two separate pre-trained DL models, InceptionV3
and DenseNet121. By fusing the results of each model using the
weighted sum rule, our system achieves exceptional accuracy
rates. Specifically, we achieve a 92.27\% detection accuracy rate,
92.33\% sensitivity, 92.22\% specificity, 90.81\% precision, and
91.57\% F1-score, outperforming existing models and demonstrating the robustness and trustworthiness of our hybrid approach.
Our study represents a significant advance in skin cancer diagnosis and provides a promising foundation for further research in
the field. With the potential to save countless lives through earlier
detection, our hybrid deep-learning approach is a game-changer
in the fight against skin cancer.}


\maketitle

\section{Introduction}\label{sec1}

Cancer happens to be one of the most lethal diseases that may occur in humans. It is the world's second leading cause of death \cite{khatun2023cancer}. There are many different kinds of cancer in the human body, and skin cancer is one of the most frequent, lethal, and rapidly spreading malignancies that can lead to death. Skin cancer develops from abnormal growth of cells that make up skin, which shields the body from heat, light and infection \cite{ashraf2020region}. It is caused by factors such as genetic factors, alcohol consumption, smoking, infections, allergies, viruses, environmental change, physical exercise,  exposure to ultraviolet light, and so on. Fungal growth and bacteria are the most likely reasons. In addition, Skin cancer is also caused by unusual swellings of the human body \cite{ali2021enhanced}.

There are various kinds of skin cancer, and among them, melanoma is the most widespread and deadliest type of skin cancer. Melanoma evolves from maliciously transformed skin melanocytes. Melanoma cells are known to spread to other areas of the body, including the liver, lungs, spleen, or brain \cite{ashraf2020region, ali2022multiclass}. Melanocytes are responsible for the production of dark pigments on the hair, skin, eyes, and different areas of the body. Melanoma lesions tend to be black or brown in color. But in a few instances, it is also found in different colors  such as azure, royal purple, rosy pink,  or even colorless \cite{zhang2020skin}.

According to the World Health Organization (WHO), cancer is one of the leading causes of death worldwide and is responsible for about 10 million deaths annually \cite{talukder2022machine}.
%add reference
It is responsible for about 1 in every 6 deaths worldwide. Between 2008 and 2030, global cancer-related deaths are expected to go up by 45\% \cite{keerthana2023hybrid}.  According to the WHO, one out of every three people in the globe has skin cancer. Melanoma and non-melanoma are the most typical cancers about 300,000 new cases were found in 2018. Melanoma cost the lives of an estimated 2490 females and 4740 males in 2019\cite{zhang2020skin}.

Deep learning (DL) is the method by which a computer model directly learns to perform categorization tasks from images, text, or sound \cite{talukder2024empowering, rasa2024brain, talukder2024toward, islam2024brainnet}. Multi-layered neural network architectures and labeled data are used to train models. DL lessens the requirement for some of the data preparation that machine learning generally demands \cite{uddin2024deep, rana2023robust}. Unstructured data like text and images can be processed by it. Gradient descent and backpropagation are used to update and tune the DL algorithm for accuracy, enabling it to make predictions with greater precision. DL capabilities have had a significant positive impact on the healthcare sector. It has several uses in various areas of medical imaging because of its high level of precision. By using these techniques, the diagnosing procedure is expedited and human error is decreased.

The most cutting-edge methods are without a doubt DL algorithms, which have lately achieved remarkable classification and segmentation results not only for medical images but also for others applications \cite{kora2022transfer, talukder2023dependable}. So, the primary reason to create the proposed hybrid DL system is to use it as a diagnostic instrument to assist individuals who make decisions in health and medical facilities in quickly and accurately identifying melanoma in dermatological images. This reduces the burden on pathologists and hospitals at a time when the number of individuals infected with skin cancer is increasing dramatically. The following are the primary contributions of this study:

\begin{itemize}
\item We proposed a hybrid DL model that can accurately classify skin cancer even in its early stages utilizing the weighted sum rule at the score-level.
\item The proposed hybrid framework has been developed by fusing the results of two different DL methods (e.g., InceptionV3 and Densenet121) that automatically extract features from the dataset and correctly detect skin cancer.
\item With a large dataset, our suggested method outperforms other existing DL models in terms of accuracy, creating a reliable and practical system that can be applied in a clinical environment to provide quick diagnoses and effective treatments.

\end{itemize}
\section{Literature Review}
\label{literature}
Dermatological photos are used to identify skin cancer. This section provides a quick overview of some recent studies on the application of DL techniques to the study of medical data. 

In order to detect skin cancer, \textit{Yuvika Gautam et al. }\cite{gautam2024fusionexnet} presented FusionEXNet, a novel fused deep-learning model. In order to take use of their capacities for feature extraction from dermoscopic images, the model combines two architectures: XceptionNet and EfficientNetV2S. Approximately 10,000 high-resolution photos from seven different kinds of skin lesions are included in the HAM10000 dataset, which was used for training and evaluation. SmoothGrad and Faster Score-CAM, two Explainable Artificial Intelligence (XAI) approaches, were integrated into the model to improve interpretability and provide insights into the model's decision-making process. FusionEXNet demonstrated superior performance in accuracy, achieving 90.83\%, compared to XceptionNet (88.82\%) and EfficientNetV2S (88.01\%). 

In order to overcome the problems of light reflections, colour fluctuations, and lesion diversity, \textit{Md Shahin Ali et al. }\cite{ali2021enhanced} suggested a deep convolutional neural network (DCNN) model for identifying benign and malignant skin lesions from dermoscopic pictures. Using filters, preprocessing eliminates noise and artefacts from the input images, normalises them, and applies data augmentation to grow the dataset size and enhance classification accuracy. The model's performance is contrasted with that of various other transfer learning models, such as MobileNet, AlexNet, ResNet, VGG-16, and DenseNet. The suggested DCNN model outperformed the other models in terms of robustness and reliability using the HAM10000 dataset, with testing accuracy of 91.93\% and training accuracy of 93.16\%.

\textit{Rehan Ashraf et al. } \cite{ashraf2020region} suggested an intelligent Region of Interest (ROI)-based melanoma detection system to overcome the difficulties caused by nevi's difficult-to-see visual appearance and deep learning models' restricted data availability. The methodology ensures that only pertinent melanoma features are incorporated for system training by using an updated k-means algorithm to extract ROIs from dermoscopic pictures. This ROI-based method concentrates on the areas that include melanoma cells, which improves feature discrimination. Subsequently, a Convolutional Neural Network (CNN) transfer learning model is trained on the ROI pictures from the DermIS and DermQuest datasets using data augmentation approaches. The suggested system outperforms current techniques that depend on comprehensive picture categorisation in terms of classification accuracy.

In this study \cite{mukadam2023skin}, dermoscopy images from the HAM10000 dataset were used to create a proprietary Convolutional Neural Network (CNN) model for skin cancer classification. The seven categories bearing and malignant that the CNN model was built to identify from the photos are essential for early detection and lowering the death rate from skin cancer. An Enhanced Super Resolution Generative Adversarial Network (ESRGAN) was used to preprocess the dataset in order to improve image quality by increasing the resolution of smaller sized images.

In this study \cite{abdar2021uncertainty}, a novel hybrid dynamic Bayesian Deep Learning  model is introduced, utilizing Three-Way Decision (TWD) theory to adaptively apply different UQ methods and neural networks in distinct classification phases. The model's effectiveness is validated on two skin cancer datasets, achieving: accuracy of 88.95\% and F1-score of 89.00\% for first dataset and accuracy of 90.96\% and F1-score of 91.00\% for 
second dataset.

This study \cite{attique2022two} uses dermoscopic images to offer a two-stream deep neural network information fusion framework for multiclass skin cancer classification. A pretrained DenseNet201 model is fed with the improved image quality through a fusion based contrast enhancement technique in the first stream of the methodology, which extracts features. An approach called skewness controlled mothflame optimisation is used to optimise these properties. A refined MobileNetV2 model's features are taken out and further downsampled in the second stream through the use of a feature selection framework. A new parallel multimax coefficient correlation technique is employed to fuse discriminant characteristics from both networks. To classify lesion images, a multiclass extreme learning machine classifier is used. Three unbalanced skin datasets HAM10000, ISBI2018, and ISIC2019, respectively are used to evaluate the concept. 

This article \cite{keerthana2023hybrid} introduces two novel hybrid CNN models that incorporate a support vector machine (SVM) classifier at the output layer. The approach involves: Extracting features from both CNN models. Concatenating these features and inputting them into the SVM for final classification.The proposed models outperformed existing state-of-the-art CNN models achieving accuracies of 88.02\% and 87.43\%. 

This study \cite{jinnai2020development} compared the diagnostic ability of dermatologists with a Faster Region-based Convolutional Neural Network (FRCNN) trained to categorise clinical photos of pigmented skin lesions. The dataset included 3551 patients' 5846 photos, with lesions classified into six classes: benign hematoma/hemangioma, nevus, seborrhoeic keratosis, senile lentigo, and malignant melanoma. A test set including 666 photos was generated, and the remaining 4732 images were annotated with bounding boxes to form the training set. In a six-class classification, the FRCNN model outperformed doctors with board certification (79.5\%) and dermatology trainees (75.1\%), with an accuracy of 86.2\%. FRCNN achieved 91.5\% accuracy in binary classification (benign vs. malignant), with 83.3\% sensitivity and 94.5\% specificity, in that order. In comparison to dermatologists, it also showed a lower false positive rate of 5.5\%. The results show that the FRCNN model performed better in picture classification tasks than 20 dermatologists, indicating that the public may benefit from using it in the future to improve skin cancer prognosis.

The goal of this work \cite{kadampur2020skin} is to create deep learning models for the classification of dermal cell pictures in order to detect skin cancer. Deep learning techniques are used at the heart of the system to improve prediction accuracy through the use of a model-driven architecture hosted in the cloud. Building these models and using them to categorise dermal cells constitute the process; standard datasets are used for testing.  With this method, practitioners may quickly develop and use deep learning models for early skin cancer identification.

This work \cite{ali2022multiclass} addressed the difficulties caused by the fine-grained heterogeneity in diagnostic categories by developing a preprocessing pipeline for multiclass skin cancer classification. Image hair removal, dataset enhancement, and image resizing to accommodate various models were all part of the pipeline. To fine-tune the EfficientNet B0-B7 models for classification, transfer learning was used on pre-trained ImageNet weights using the HAM10000 dataset. With an F1 Score of 87\% and Top-1 Accuracy of 87.91\%, EfficientNet B4 had the best results when the models' performance was assessed using Precision, Recall, Accuracy, F1 Score, and Confusion Matrices. More complicated models do not always perform better; models with intermediate complexity, such EfficientNet B4 and B5, showed notable superiority over more complex models.

This research \cite{javid2023design} utilizes images from various publicly available ISIC  datasets to create a balanced dataset consisting of 10,500 images for training and testing. An ensemble of four convolutional neural network (CNN) architectures-ResNet50, EfficientNet B6, InceptionV3, and Xception was employed for melanoma classification.The experimental results demonstrate that the proposed ensemble model effectively classifies melanoma skin cancer with high accuracy, outperforming many state-of-the-art methods.

\section{Methodology}

\begin{figure}[htbp]
\centering
\includegraphics[scale=.7]{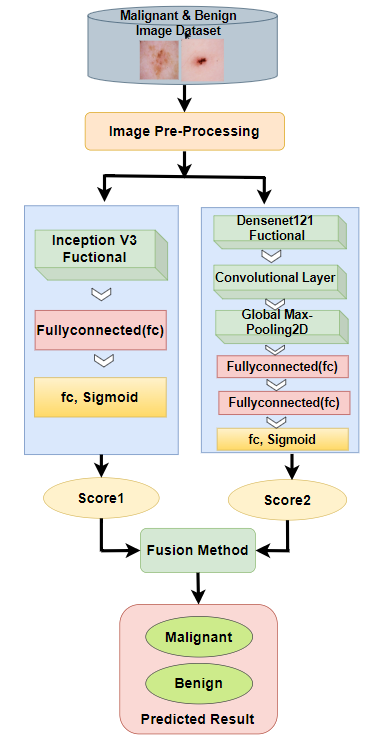}
\caption{The block diagram of the Hybrid model for Skin cancer detection.}
\label{fig:proposed}
\label{flow}
\end{figure}

The main principle of our idea is depicted in the block diagram in figure \ref{flow}.
% use figure reference and describe the methodology in brief. And where is the figure?
This section presents the framework, architecture, classification and performance evaluation of a hybrid DL model for detecting skin cancer.

\subsection{Dataset}\label{AA}
Our dataset basically consists of two distinct categories of skin carcinoma images. DL requires a significant amount of data in order to produce a good result. The collection of skin cancer pictures, on the other hand, is critical. To address these issues, we used a freely available dataset of 3297 dermoscopic images from Kaggle \cite{ Skin}. It includes 1800 images of benign and 1497 images of malignant categorized moles. The data was obtained via ISIC Archive. Each picture is 224x244 pixels in size. The professional pathologist confirmed the database's ground truth.

To prevent overfitting, the dataset is split between 80\% training data and 20\% testing data that results in 2637 training images and 660 test images out of the total number of images. 10\% of the training set was arbitrarily selected during the learning process and used as  a validation set to criterion the generalizability of the Model and save the weights set that produces the lowest level of an error on the validation collection.
\subsection{Image Preprocessing}\label{BB}
The main goal of the image preprocessing process is to make the data suitable for deep learning models and improve the initial medical images by getting rid of noise, air bubbles, and artifacts caused by gel that was sprayed before the picture was taken. To attain a high classifying rate, we removed artifacts and noise from the pictures. The elimination of this noise and artefacts was crucial to guaranteeing high-quality input images since they could mask important features required for classification. The image was smoothed using noise reduction algorithms, which also removed any random deviations that could confound the model. We employed feature scaling and turned the categorical label data into number data\cite{talukder2023dependable}. In order to ensure that the input data had consistent numerical ranges, we used feature scaling to normalise the pixel intensity values, which can vary significantly in photos. This step prevents gradient instability problems and helps models converge more quickly during training.

\subsubsection{Class Labeling}
We used label encoding techniques to transform the categorical data into numerical format because the labels in medical picture datasets are typically categorical (e.g., different forms of skin cancer). In order to enable the deep learning model to process and learn from the labels during training, each distinct category was given a numerical value.
 The labeling data collection is presented as follows: the image's label 0 indicates benign patients, while 1 symbolizes malignant patients\cite{ahmad2022novel}. % why citation is used here?

\subsubsection{ Feature scaling}

The feature scaling method is well recognized in the field of machine learning and pattern recognition for being used to normalize data. To avoid outliers, this approach sets all data items to the same scale and thus improves prediction quality. As the features in cancer datasets have a high variance,  One of the pre-processing strategies for normalization is feature scaling.
%Rewritre the sentence more precisely
We divided the grayscale value of an image by 255 to standardize our image pixels between 0 and 1. As a result, the numbers will be relatively little, and the computation will be simpler and faster\cite{basavegowda2020deep,sun2020new}. % why citation is used here?

\subsection{Transfer Learning}

Transfer Learning has emerged as a popular method in Deep Learning (DL) that utilizes a pre-trained model for a similar task to train a new model. This approach is particularly useful in medical imaging, where training on large datasets such as ImageNet with all neural network parameters can be computationally intensive \cite{talukder2022machine}. The ImageNet dataset, for instance, contains 14 million images with 20,000 categories for visual recognition tasks \cite{hossain2022automatic}.

In this study, we propose a model for skin lesion classification, which was trained and evaluated using advanced Convolutional Neural Networks (CNNs) such as Inception V3, Densenet121, and a Hybrid model \cite{kadampur2020skin,ali2021enhanced}. To optimize our model, we replaced the last softmax layer with a sigmoid activation, reducing the classification output to two classes for identifying benign and malignant skin lesions.

The Inception V3 CNN model achieved remarkable accuracy of over 78.1\% on the ImageNet dataset and has been widely used for image classification tasks \cite{basavegowda2020deep}. Its architecture includes inception modules that use filters of different sizes on the same input, and auxiliary classifiers for regularization weight loss ratio. The convolution layer implements factorization to minimize dimensionality and avoid overfitting.

Another CNN architecture utilized in our model is DenseNet121, which resolves the vanishing gradient problem by directly connecting every layer to every other layer. Instead of summing feature maps from previous layers, DenseNet121 concatenates them, resulting in a compact model with fewer parameters and enabling feature reuse \cite{huang2017densely}.

Overall, our proposed model incorporating Inception V3, Densenet121, and a Hybrid model shows promising results for skin lesion classification and has the potential to improve diagnostic accuracy in dermatology.

\subsection{Proposed Model}
After the preparation stages, the following section involves feeding data into the suggested model. This section introduces the Inceptionv3, Densenet121 and Hybrid model for skin cancer detection and classication.

\subsubsection{Inception V3}
InceptionV3 is a CNN-based DL model used for image classification. It is a highly renowned image detection model that has been developed over time by a number of researchers. On the ImageNet dataset, the model achieved more than 78.1\% accuracy. Google's InceptionV3 is the third version in the DL Convolutional Architecture series. The model was developed based on 1000 classes from the ImageNet dataset, which was trained on approximately 1 million images. It is made up of inception modules that apply filters of various sizes to the same amount of input. In the Inception network, the auxiliary classifiers add to the regularization weight loss ratio. Factorization was implemented in the convolution layer in order to minimize dimensionality and therefore overfitting \cite{basavegowda2020deep}. 
 \begin{figure}[h]
\centerline{\includegraphics[width=.5\textwidth]{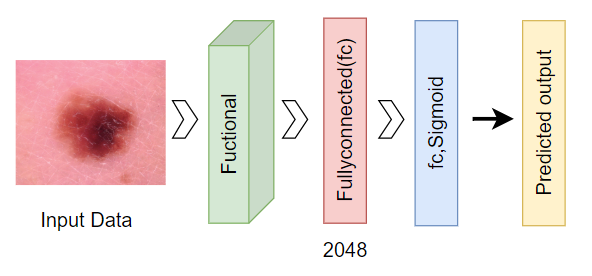}}
 \caption{InceptionV3 network structure of proposed model.}
\label{fig: incp2}
\end{figure}

In this study, the weights of the pre-trained InceptionV3 model were used as an initial step, followed by two fullyconnected layers. Then, It was fine-tuned using Adam optimizer over the current training set. The main idea is that the pre-trained InceptionV3 model has extensive knowledge of detecting various types of feature representations (e.g., edges, curves, corners, etc.), and that fine-tuning its parameters will allow the InceptionV3 model to quickly learn the specific feature representations of the current task ,as shown in figure \ref{fig: incp2}.  

\subsubsection{Densenet121}
\begin{figure}[htbp]
 \centerline{\includegraphics[width=.8\textwidth]{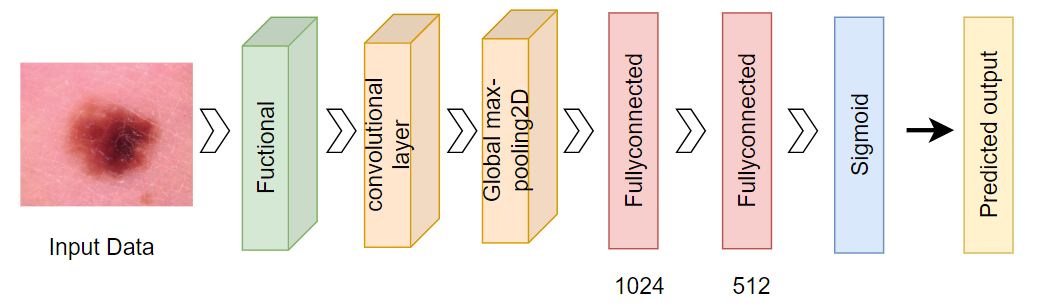}}
 \caption{Densenet121 network structure of proposed model.}
 \label{fig: dense}
 \end{figure}

DenseNets resolve the vanishing gradient problem arising from traditional CNN. The term "Densely Connected Convolutional Network" refers to an architecture in which every layer is directly connected to every other layer. Instead of summing the feature maps from the prior layers, they are concatenated and utilized as inputs in each layer. Consequently, DenseNets need less parameters than typical CNNs, allowing for the removal of duplicate feature maps and feature reuse. DenseNets are divided into DenseBlocks,  where the number of filters between them changes but the size of the feature map within a block stays the same. DenseNets produce more compact models and achieve state-of-the-art performance because they require fewer parameters and enable feature reuse.
In this study,  weights of the pre-trained Densenet121 model the were used as an initial step, followed by one convolutional layer and one global max-pooling layer. Then, three fully connected layers were used and It was fine-tuned using Adam optimizer over the current training set. The main idea is that the pre-trained Densenet121 model has extensive knowledge of detecting various types of feature representations (e.g., edges, curves, corners, etc.), and that fine-tuning its parameters will allow the Densenet121 model to quickly learn the specific feature representations of the current task, as shown in figure \ref{fig: dense}.
\subsection{Feature Fusion Technique}

Feature fusion methods are computer programs that use scores from different models to make a decision.  A single scalar score is often generated as a result of this consolidation process. It typically utilizes relatively easy fusion operators and doesn't require a lot of computing. It has been found that using outputs with similar performances when combining different score-level fusion algorithms results best. A reasonably simple score-level fusion technique that works directly with raw score data is the sum rule. Sum rule fusion can be either a simple sum rule or a weighted sum rule. The weighted sum rule is shown as follows:
\begin{equation}
     f_s = {w_1x_1 + w_2x_2+ ... +w_nx_n}
 \end{equation}
Where \( w_{\text{i}}\) is the weight value assigned to the score value of the $i$th model.
According to the sensitivity analysis, the sum rule is the most durable method for error estimation. It has been found that the Sum rule performs better than most other state-of-the-art score-level fusion algorithms\cite{ezichibiometric}.

\subsection{Feature Fusion Hybrid Model}

In this study, we proposed the use of Densenet121 and InceptionV3 models for the accurate classification of skin cancer. The previously trained InceptionV3 model was utilized by employing its weighted values as an initial phase, which was then followed by two fully connected layers. Subsequently, the model was fine-tuned using the Adam optimizer on the present training set. The rationale behind this approach is based on the premise that the pre-trained InceptionV3 architecture has a deep understanding of detecting various feature representations, and by tuning its parameters, the model can quickly discover the particular feature representations of the given task.

Similarly, for the Densenet121 model, we utilized the pre-trained weights as an initial step, which was then followed by one convolutional layer and one global max-pooling layer. To further enhance the model's classification accuracy, we employed three fully connected layers, which were fine-tuned on the present training set using the Adam optimizer. The key insight behind this approach is that the pre-trained Densenet121 model has in-depth expertise in identifying various kinds of feature representations, and by changing its parameters, the model can swiftly learn the particular feature representations of its present task.

In this paper, we introduced a feature fusion hybrid model for improved performance and novel technique for skin cancer classification. Our proposed system combines the output from two powerful models, InceptionV3 and Densenet121, to make the final decision. For each input image, two anticipated probability ratings are generated, and the input image is classified into either the benign or melanoma class based on the highest probability score.

To combine the results from two models, we used the weighted sum rule, taking into account the parallel architecture in the suggested system. This approach provides dermatologists with a high level of confidence to make the final judgment and accurately discriminate between cancer-infected and healthy patients. During the model implementation, we assigned a slightly higher weight value to the Densenet121 model than the InceptionV3 model, as it outperformed the latter in our experiments.

Our proposed hybrid framework significantly improves the accuracy of skin cancer classification when compared to using either the InceptionV3 or Densenet121 model alone. With our feature fusion approach, we aim to provide a powerful tool for dermatologists to improve their diagnoses and ultimately improve patient outcomes.

Overall, our study demonstrates that leveraging pre-trained deep learning models such as Densenet121 and InceptionV3 can significantly improve the accuracy of skin cancer classification. By fine-tuning these models on the present training set, we were able to achieve state-of-the-art performance, highlighting the potential of deep learning in the field of medical image analysis.
Table \ref{tab:tuning} highlights the parameters that are amended in our proposed hybrid model.

% \section{Training and performance}

% TABLE 

% . The primary ones we utilized in our experiment are listed below-
% \begin{itemize} 
% \item Pooling layer (MaxPooling2D): Is employed to shrink the size of the input pictures for quicker training.
% \item Batch size (32): Number of processed images in every iteration.
% \item Learning rate (0.0001): Specifies the rate at which the learning process will start.
% \item Optimizer (Adam): In order to reduce the loss function when training DL models, Adam is a stochastic gradient descent replacement optimization approach. 
% \item Loss function (sparse categorical cross-entropy): This is done in order to construct an issue of binary classification from the dataset.
% \item Number of epochs (50): The number of complete passes through the training dataset.
% \item Number of class (2): Melanoma and benign.

% \end{itemize}

% \usepackage{graphicx}
\begin{table}[h]
\centering
\caption{Hyperparameters used in proposed models}
% \label{tab:}
\begin{tabular}{p{4cm}p{8.5cm}}
\hline
\textbf{Hyperparameter} & \textbf{Description} \\ \hline                                                 
MaxPooling2D layer & Shrink input pictures for faster learning. \\ Batch size & 32 (photos processed per iteration) \\ 
Learning rate & 0.0001 (rate of learning process start) \\ 
Optimizer & Adam (stochastic gradient descent replacement) \\ 
Loss function & Sparse categorical cross-entropy (binary classification) \\ 
Epochs & 50 (number of passes through training dataset) \\ 
Classes &  (2) Melanoma and benign \\ \hline
\end{tabular}%
\label{tab:tuning}
\end{table}
\section{Result }

\subsection{Environment Setup}
 The model was implemented in Kaggle  on a system with Intel Core i7-1165G7 CPU processor, 8 GB of RAM, and NVIDIA GeForce MX330 GPU card.
 \subsection{Evaluation Metrics}
 The effectiveness of the suggested model is assessed by taking into account the evaluation metrics based on Accuracy Score, Precision, Recall, F1-measure, Specificity, confusion matrix and ROC curve. 
 \begin{itemize}

\item \textbf{Confusion Matrix: } As illustrated in fig \ref{fig:cm}, the confusion matrix is a technique for assessing performance in the form of a table that incorporates information about both actual and expected classes. The dimension of the confusion matrix would be nxn if the proposed problem to be examined has an n row, with the rows representing the actual row and the columns representing the predicted row. For two or more classes, the matrix depicts actual and anticipated values.
 
 \begin{figure}[htbp]
    \centering
    \includegraphics[scale=0.2]{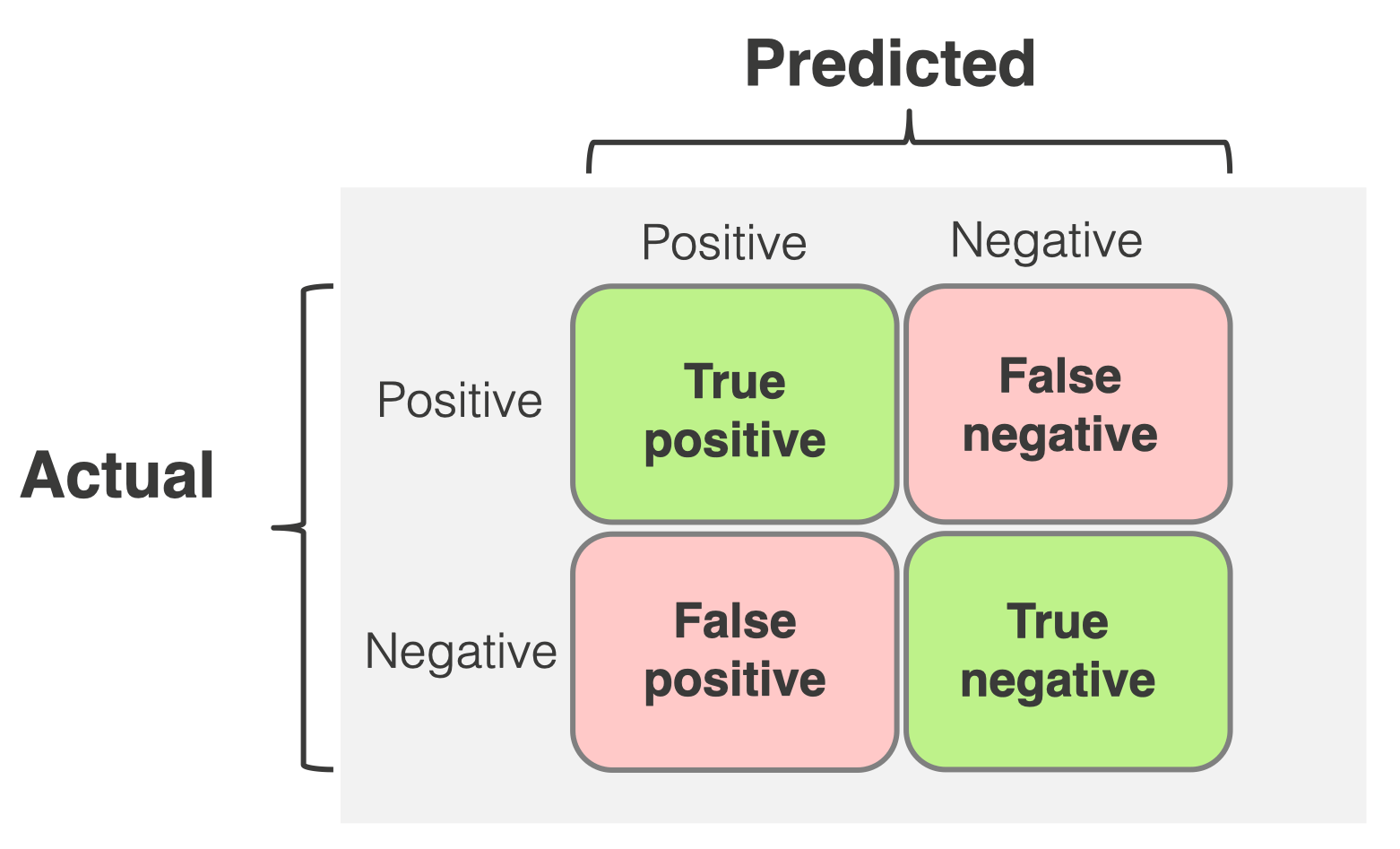}
    \caption{The Confusion Matrix}
    \label{fig:cm}
\end{figure}

The explanation of the terms
of TP, FP, TN, FN are-

  \begin{itemize}
    \item The total number of correct outcomes or forecasts where the actual class was positive is known as true positive (TP).
    \item  The total number of incorrect results or forecasts while the actual class was positive is known as false positive (FP).
    \item The total number of correct outcomes or predictions where the actual class was negative is known as true negative (TN).
    \item The total number of incorrect outcomes or forecasts when the actual class was negative is known as false negative (FN).
    \end{itemize}

     \item \textbf{Accuracy Score:}
Accuracy measures how properly the classifier anticipates the classes \cite{fakoor2013using}. The average number of samples is correctly classified by the classifier. It is mathematically defined as:

\begin{equation}
 Accuracy = \frac{Correctly\; Predicted\; Data}{Total\; Testing\; Data} \times 100\% = \frac{TP+TN}{TP+TN+FP+FN}\times 100\%
\end{equation}

\item \textbf{Precision: }
Precision is the proportion of all the model's positive classifications that are actually positive. It is mathematically defined as:

\begin{equation}
     Precision = \frac{Correctly\;Classfied\;Actual\;Positive}{Everything \;Classified \;as \;Positive}= \frac{TP}{TP+FP}
\end{equation}

\item \textbf{Recall:}
Recall is also known as true Sensitivity, which is the percentage of all real positives that were appropriately identified as positives. It is mathematically defined as:

\begin{equation}
      Recall = \frac{Correctly\;Classified\;Actual\;Positive}{All\;Actual\;Positives}=\frac{TP}{FN + TP}
 \end{equation} 

 \item \textbf{F1 measure : }
 The harmonic mean of precision and recall is F1. For the best classification results, it should be close to one, while for the worst classifiers, it should be close to zero. It is mathematically defined as:
 
\begin{equation}
    F_1 = 2 \times \frac{Precision \times Recall}{Precision + Recall}
\end{equation} 

\item \textbf{Specificity: } Specificity is the metric that evaluates a model’s ability to predict true negatives of each available category. These metrics apply to any categorical model. The equations for calculating this metrics are as follows-

\begin{equation}
      Recall =\frac{TN}{FP + TN}
 \end{equation} 

 \item \textbf{Receiver Operating Characteristic curve: } The Receiver Operating Characteristic curve, or the ROC curve, is a valuable tool for forecasting the likelihood of a binary result. It's a graph that shows the false positive rate (x-axis) against the true positive rate (y-axis) for a variety of options. When the actual outcome is positive, the true positive describes how well the model predicts the positive class. The curve's shape offers a wealth of information, including what we would care about most for an issue, the expected false positive rate, and the expected false-negative rate. To be precise about this:
 \begin{itemize}
    \item Lower false positives and higher true negatives are shown by lower values on the x-axis of the plot.
    \item  Greater true positives and smaller false negatives are shown by higher values on the y-axis of the graphic.

 \end{itemize}

 \end{itemize}

\subsection{Result analysis}

The proposed hybrid model combines the InceptionV3 and DenseNet121 models. Table \ref{tab:performance} reports the quantitative performance evaluation of three deep learning models, namely InceptionV3, Densenet121, and Hybrid, in terms of precision, sensitivity, specificity, F1-score, and accuracy. The results indicate that the Hybrid model outperforms the other two models, achieving an accuracy of 92.27\%, which is 1.67\% and 1.67\% higher than InceptionV3 (90\%) and Densenet121 (90.6\%), respectively. The precision and sensitivity of all models are relatively high, with values above 89\%. Specifically, the precision of InceptionV3, Densenet121, and Hybrid is 89.6\%, 89.7\%, and 90.8\%, respectively, while their sensitivity is 89.6\%, 90\%, and 92.33\%, respectively. The Hybrid model also achieves the highest F1-score of 91.57\% among the three models, with InceptionV3 and Densenet121 obtaining F1-scores of 89.6\% and 90.3\%, respectively. The specificity of all models is around 91.3\%, with Hybrid achieving a slightly higher value of 92.22\%. Overall, the Hybrid model, fuse the results from three deep learning models using the weighted sum rule, demonstrates superior performance compared to the other two models in terms of accuracy, F1-score, and sensitivity.

\begin{table}[h]
\centering
\caption{Performance analysis of our models}
\label{tab:performance}
\begin{tabular}{p{2cm}p{2cm}p{2cm}p{1.5cm}p{1.5cm}p{1.5cm}}
\hline
Model Name            & Precision & Sensitivity & Specificity & F1\_score & Accuracy \\ \hline
InceptionV3           & 89.6      & 89.6        & 91.3        & 89.6      & 90       \\
Densenet121           & 89.7      & 90          & 91.3        & 90.3      & 90.6     \\
Hybrid  & 90.80     & 92.33       & 92.22       & 91.57     & 92.27    \\ \hline
\end{tabular}%
\end{table}

\begin{figure}[htbp]
    \centering
    \subfigure[Accuracy graph for Densenet121]{\includegraphics[scale=0.6]{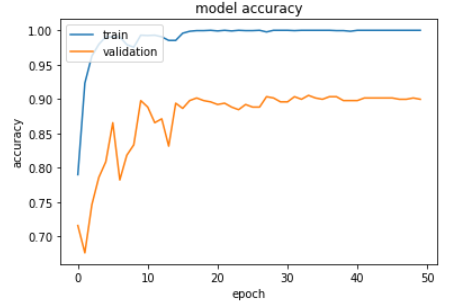}
    \label{fig:con10}
    }
     \subfigure[Loss graph for Densenet121
]{\includegraphics[scale=0.6]{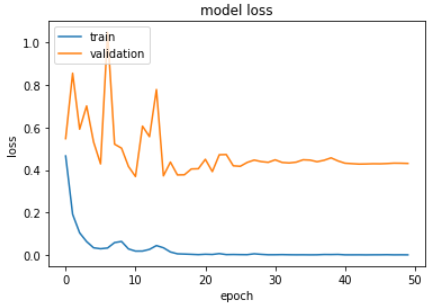}
    \label{fig:con20}
    }
    \subfigure[Accuracy graph for InceptionV3
]{\includegraphics[scale=0.6]{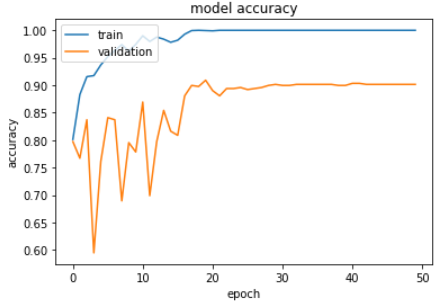}
    \label{fig:con30}
    }
    \subfigure[Loss graph for InceptionV3
]{\includegraphics[scale=0.6]{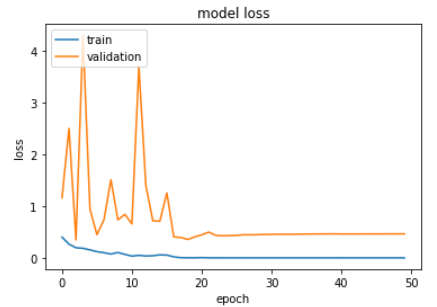}
    \label{fig:con40}
    }
    \caption{Accuracy and Loss graph of Densenet121 and InceptionV3}
    \label{fig:enter-label}
\end{figure}

Figures \ref{fig:con10}, \ref{fig:con20}, \ref{fig:con30} and \ref{fig:con40} display the performance curves (accuracy and loss) of Densenet121 and InceptionV3 deep learning models during training. These curves depict changes in accuracy and loss with increasing epochs. Performance curves are crucial in assessing model effectiveness in classification tasks, aiding in identifying overfitting or underfitting, and gauging the model's performance improvement over time.

\begin{figure}[htbp]
\centerline{\includegraphics[width=.5\textwidth]{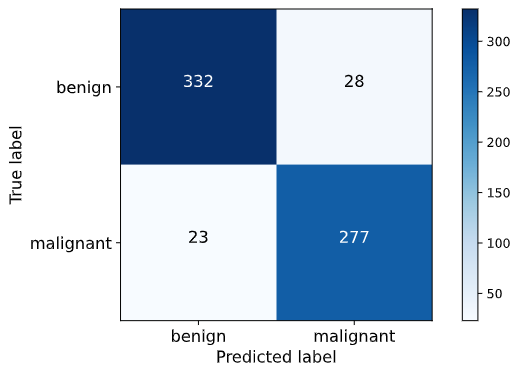}}
\caption{Confusion Matrix of testing data.}
\label{fig:confusion_matrix}
\end{figure}

The confusion matrix for our proposed skin cancer classification model is presented in Fig \ref{fig:confusion_matrix}. The rows correspond to the actual classes, while the columns represent the predicted classes. The cells of the matrix show the number of instances that were classified in each class. In our proposed model, 332 instances of benign skin cancer were correctly classified as benign (true negatives), while 28 instances of benign skin cancer were wrongly classified as malignant (false positives). Similarly, 27 instances of malignant skin cancer were correctly classified as malignant (true positives), while 23 instances of malignant skin cancer were wrongly classified as benign (false negatives). The results show that the proposed system shows great potential to reduce the workload of dermatologists and assist in correctly identifying malignant skin lesions in clinical practice.

\begin{figure}[htbp]
\centerline{\includegraphics[width=.6\textwidth]{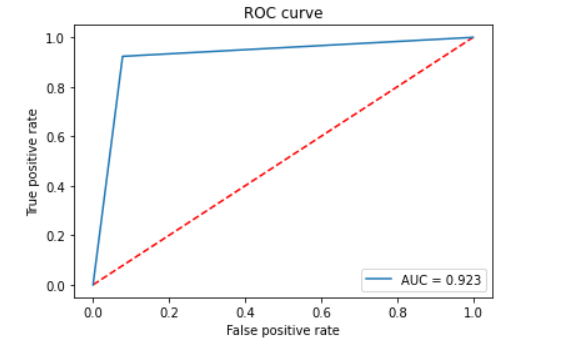}}
\caption{Receiver Operating Characteristic curve of our proposed Hybrid model.}
\label{fig:con50}
\end{figure}

Fig.\ref{fig:con50} displays the Receiver Operating Characteristic (ROC) curve, which was constructed by calculating the true positive ratio and false positive ratio for various accuracy thresholds. It is evident from the graph that the Hybrid model proposed in this study has achieved an impressive area under the ROC curve of 92.3\%.

\section{Discussion}

Table \ref{tab:comparision1} presents a comparison of five state-of-the-art models proposed for skin cancer detection by different authors. M. S. Ali et al. \cite{ali2021enhanced} proposed a deep convolutional neural network (DCNN) model, achieving an accuracy of 91.93\% on the HAM10000 dataset. K. Ali et al. \cite{ali2022multiclass} proposed an EfficientNetB4-based model, achieving an accuracy of 87.91\% on the same dataset. D. Keerthan et al. \cite{keerthana2023hybrid} proposed an SVM-based model, achieving an accuracy of 88.2\% on the ISBI2016 dataset. M. Abdar et al. \cite{abdar2021uncertainty} proposed a two-way deep belief network with deep learning (TWDBDL) model, achieving an accuracy of 88.95\% on the Kaggle dataset.

In comparison, the model proposed in this study, which is a hybrid approach, achieved the highest accuracy of 92.27\% on the ISIC dataset, outperforming all other models. This model holds promise for clinical application in skin cancer detection. Moreover, the performance of the recommended is sufficient for the diagnosis of melanoma under various environmental scenarios.

\begin{table}[htbp]
\caption{Comparison of Proposed Model.}
\label{tab:comparision1}
\begin{tabular}{p{3.5cm} p{1cm} p{3cm} p{1.5cm} p{2.5cm}}
\hline
\textbf{Authors}   & \textbf{Year}   & \textbf{Methodology}      &\textbf{Dataset} &  \textbf{Accuracy} \\ \hline

M. S. Ali et al. \cite{ali2021enhanced}   & 2021      & DCNN   & HAM10000  & 91.93\%     \\ 

 K. Ali et al. \cite{ali2022multiclass}   & 2022      &  EfficientNetB4    & HAM10000  & 87.91\%    \\
 
 D. Keerthan et al.  \cite{keerthana2023hybrid}   & 2023      & SVM    & ISBI2016  & 88.2\%     \\ 

 M. Abdar et al.  \cite{abdar2021uncertainty}   & 2021      & TWDBDL    & Kaggle  & 88.95\%   \\ 
 
 H.-Y. Huang et al. \cite{huang2023classification}   & 2023      & YOLOv5    & ISIC  & 79.2\%    \\ 
  Yuvika Gautam et al. \cite{gautam2024fusionexnet}   & 2024      & FusionEXNet    & HAM10000  & 90.83\%   \\ 
   Hatice Catal Reis et al. \cite{reis2024fusion}   & 2024      & MABSCNET \& ViT    & ISIC  & 78.63\%  \& 76.50\% \\ 

\textbf{Proposed Model} &    \textbf{2024 }    & \textbf{Hybrid}  & \textbf{ISIC} & \textbf{92.27\%  }   \\ \hline

\end{tabular}
\end{table}

\section{Conclusions}
This research looked into the feasibility of developing a hybrid system for skin cancer detection that uses deep learning techniques to provide accurate and real-time diagnostics. Data preparation enhanced the image quality and reduced noise levels. Both the InceptionV3 and DenceNet121 were trained on the preprocessed images to improve the accuracy performance. By using the sigmoid for the output activation layer, the suggested technique can differentiate between malignant and benign skin lesions. Likewise, we tested the proposed approach on a dataset from Kaggle and found that it performed better for training and testing accuracy. The proposed method has managed to accomplish an acceptable level of efficiency with an accuracy of 92.27\%, precision of 90.81\%, sensitivity of 92.33\%, specificity of 92.22\%,  and F1-score of 91.57\%. This study may lessen the amount of pressure on decision-makers (such as clinicians and therapists) as a result of the growing number of cancer patients in comparison to the shortage of medical resources. In the future to achieve the most accurate prediction and classification accuracy, we will employ more advanced pre-processing methods to successfully build a model from a sizable dataset that has more identified skin lesions.

\backmatter

%\bmhead{Supplementary information}

%If your article has accompanying supplementary file/s please state so here. 

%Authors reporting data from electrophoretic gels and blots should supply the full unprocessed scans for key as part of their Supplementary information. This may be requested by the editorial team/s if it is missing.

%Please refer to Journal-level guidance for any specific requirements.

%\bmhead{Acknowledgements}

%Acknowledgements are not compulsory. Where included they should be brief. Grant or contribution numbers may be acknowledged.

%Please refer to Journal-level guidance for any specific requirements.

\section*{Declarations}

\begin{itemize}
 \item \textbf{Funding :} This research was supported by the National Science and Technology  for research fellowship of 2022-23.
\item \textbf{Conflict of interest/Competing interests:} The authors declare no conflict of interest. 
\item \textbf{Ethics approval and consent to participate:} Not Applicable
\item \textbf{Statement of human and animal rights:} On behalf of all authors, the corresponding author affirms that human and animal rights were upheld in the study.
\item \textbf{Consent statement:} On behalf of all authors, the corresponding author states that informed consent was obtained from all participants involved in the study.
\item \textbf{Data availability}: The selected datasets are sourced from free and open-access source: \url{https://www.kaggle.com/datasets/fanconic/skin-cancer-malignant-vs-benign?select=train}
\item \textbf{Materials availability: } Not Applicable
\item \textbf{Code availability: } Not Applicable
\item \textbf{Author contribution: } Conceptualization: Maksuda Akter, Rabea Khatun, Md. Manowarul Islam; Supervision: Md. Manowarul Islam; Methodology: Maksuda Akter, Rabea Khatun, Md. Manowarul Islam,;  Data curation: Maksuda Akter, Rabea Khatun; Resources: Maksuda Akter, Rabea Khatun, Md. Manowarul Islam; Software: Maksuda Akter, Rabea Khatun, Md. Manowarul Islam; Formal analysis,  Maksuda Akter, Rabea Khatun,  Md. Manowarul Islam, Md. Alamin Talukder; Writing – original draft: Maksuda Akter, Rabea Khatun; Validation, Md. Manowarul Islam, Dr. Md. Ashraf Uddin.

\end{itemize}

\bigskip
%%Editorial Policies for:

%\bigskip\noindent
%Nature Portfolio journals: %\url{https://www.nature.com/nature-research/editorial-policies}

%\bigskip\noindent
%\textit{Scientific Reports}: %\url{https://www.nature.com/srep/journal-policies/editorial-policies}

%\bigskip\noindent
%BMC journals: %\url{https://www.biomedcentral.com/getpublished/editorial-policies}
%\end{flushleft}





\bibliography{reference}% common bib file

\end{document}